\documentclass[twocolumn,showpacs,aps,prl]{revtex4}
\usepackage{graphicx}
\usepackage{dcolumn}
\usepackage{amsmath}
\usepackage{latexsym}

\begin {document}
\title {About the fastest growth of Order Parameter in Models of Percolation}
\author
{
S. S. Manna
}
\affiliation
{
\begin {tabular}{c}
Satyendra Nath Bose National Centre for Basic Sciences,
Block-JD, Sector-III, Salt Lake, Kolkata-700098, India \\
\end{tabular}
}
\begin{abstract}
Can there be a `Litmus test' for determining the nature of transition in models of percolation? In this paper we argue 
that the answer is in the affirmative. All one needs to do is to measure the `growth exponent' $\chi$ of the largest 
component at the percolation threshold; $\chi < 1$ or $\chi = 1$ determines if the transition is continuous or discontinuous. 
We show that a related exponent $\eta = 1 - \chi$ which describes how the average maximal jump sizes in the Order 
Parameter decays on increasing the system size, is the single exponent that describes the finite-size scaling of a number 
of distributions related to the fastest growth of the Order Parameter in these problems. Excellent quality scaling analysis 
are presented for the two single peak distributions corresponding to the Order Parameters at the two ends of the maximal 
jump, the bimodal distribution constructed by interpolation of these distributions and for the distribution of the maximal jump in 
the Order Parameter. 
\end{abstract}
\maketitle

      Macroscopic properties change when a thermodynamic system undergoes a phase transition from one phase 
   to another. The transition is characterized by looking at how a certain observable, called the Order Parameter 
   (OP), varies before, at and beyond the critical point of a transition while a suitable control variable is 
   continuously tuned. The OP exhibits a very rapid change in magnitude starting from the critical point. Depending 
   on if this rapid change occurs discontinuously or continuously, the transition is termed as discontinuous 
   (or first order) or continuous (or second order) transition 
   \cite {Stanley}.

      Recently it has been seen that such straight forward classification may not be easy in some models of
   percolation phenomena \cite {Stauffer,Grimmett}. Originally the percolation model was introduced by Broadbent 
   and Hammersley in 1957 \cite {Broadbent} to better understand the mechanism of charcoal based gas masks and 
   later this model had been used extensively to study the order / disorder transitions in various systems. Long 
   range correlation in terms of global connectivity appears as the density of nodes / links is increased beyond 
   certain critical point, called the percolation threshold. Over several decades many different variants of 
   percolation model had all exhibited only continuous transitions. This was the situation until recently when 
   Achlioptas et. al. argued that a local competition between a pair of vacant edges for being occupied indeed 
   leads to an abrupt jump in OP. Hence they coined the name `Explosive Percolation' (EP) \cite {ACH} to emphasize
   the abruptness of the transition.

      The original model of EP had been defined by a slight modification of Random Graphs \cite {Erdos} and we 
   refer to it as Achlioptas process (AP) in the following. One starts with $N$ nodes (each node is a component of 
   size unity) and no links. At an 
   arbitrary intermediate stage a pair of vacant edges are randomly selected between nodes $(i,j)$ and $(k,l)$ 
   which belong to the components of sizes $s_i$, $s_j$ and $s_k$, $s_l$ respectively. There is a competition: 
   if $s_is_j < s_ks_l$ then link $(i,j)$ is occupied, otherwise $(k,l)$ is occupied; however when the products 
   are equal, one of the two edges is selected randomly. This bias towards connecting small components delays 
   the growth of the largest component and consequently the percolation threshold is pushed up to $p_c= 0.888449(2)$ 
   \cite {Grassberger} compared to 1/2 for Random Graphs. In general the preferential link occupation rule slows down
   the growth of largest component and accelerates smaller components to grow faster. Further EP has been studied on 
   the Square Lattice \cite {ZIFF,ZIFF1}, Random Graph \cite {Friedman}, scale-free networks \cite {Cho,Radicchi,Radicchi1}, 
   with a Hamiltonian formulation \cite {Herrmann}, in a cluster aggregation process \cite {Cho1} and also in human 
   protein homology network \cite {Rozenfeld}, real-world networks \cite {Pan} etc. 

      Subsequently situation started changing and a number of research publications suggested that EP transition is 
   actually continuous. The first claim came from da Costa et. al. who considered a slightly modified version of AP 
   \cite {Costa}. For each of the randomly selected pair of vacant edges one first picks up the node with smaller 
   component size and then connects them. It was argued that though this model is more biased than AP it has a 
   continuous transition and therefore the original AP must then have a continuous transition. Later it was shown 
   numerically that the magnitude of the maximal jump in OP has a zero measure in the thermodynamic limit \cite 
   {Manna,Nagler}. Very recently Riordan and Warnke have rigorously proved that all models using Achlioptas type
   processes are indeed continuous in the asymptotic limit \cite {Riordan}.

      A distinction between two types of EP models can be made: (i) Local rule: As in Random Graphs these models 
   pick up in general a few nodes randomly with uniform probability and then preferentially occupy a vacant edge within 
   this subset of nodes to introduce the bias towards connecting smaller components. (ii) Global rule: The entire set
   of vacant edges are assigned non-uniform probabilities which are already biased towards smaller components and out of 
   them one vacant edge is randomly selected for occupation. In such a model a vacant edge is selected with a probability 
   proportional to $(s_is_j)^{\zeta}$ and it is argued that for all values of $\zeta < \zeta_c$ the transition is 
   discontinuous \cite {Manna}.

      The distribution of the Order Parameter at the percolation threshold has a bimodal distribution where
   the depth of valley between the two peaks vanishes as $N \to \infty$ which signifies a discontinuous transition \cite {Grassberger,
   Radicchi1,Tian}. Extensive scaling analysis of these distributions in a set of four Local rule models showed that all 
   of them have continuous transitions \cite {Grassberger}. The positions of peaks scale with $N$ with different 
   exponents which indicate different growth mechanisms for the sub and super-critical percolation regimes \cite 
   {Grassberger}.

\begin{figure}
\begin {center}
\includegraphics[width=7.0cm]{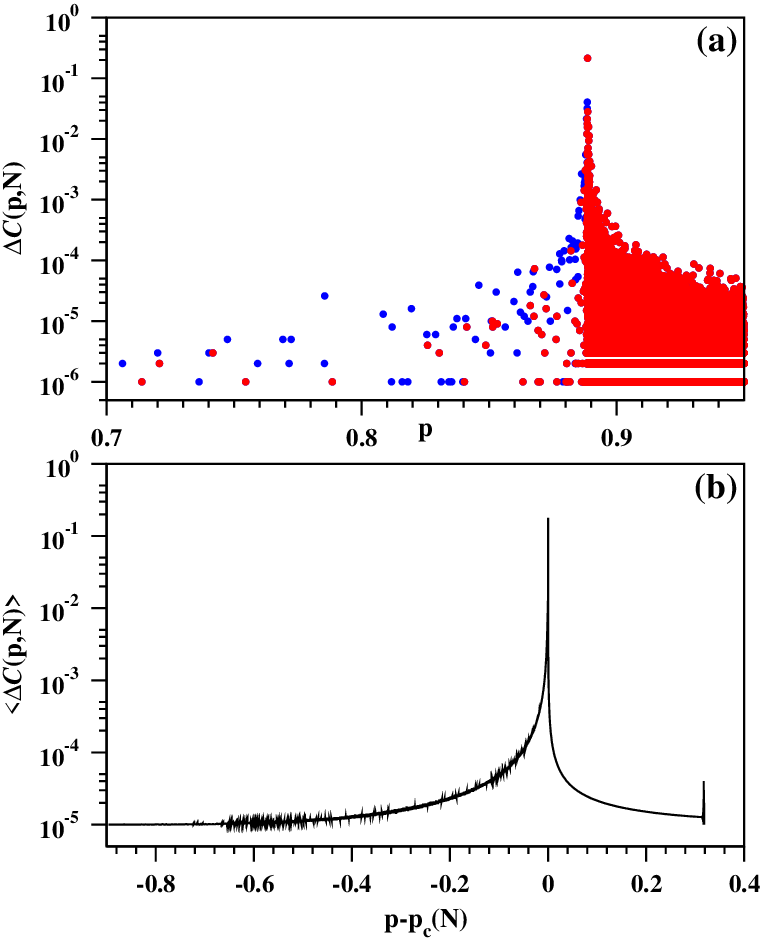}
\end {center}
\caption{(Color online) The AP data of a single run. 
(a) The jump sizes $\Delta {\cal C}(p,N)$ in the Order Parameter have been plotted (blue symbols) with link density $p$ for the 
graph of size $N=10^6$. The subset of red symbols corresponds to those jumps where existing largest components increased their sizes.
(b) The average jump size $\langle \Delta {\cal C}(p,N) \rangle$ plotted with deviation $p-p_c(N)$ from the percolation threshold for $N=10^5$. }
\end{figure}

      In all percolation models we have studied, we start from the empty graphs with $N$ nodes and no edges. Links are then dropped
   one by one following the specific rules of the model and sizes of different components of the graph start growing. The 
   link density is defined as $p=n/N$ where $n$ is the number of links in the graph. If $s_{max}$ is the size of the largest component 
   of the graph then the Order Parameter is defined as ${\cal C}(N)=s_{max}/N$. To characterize the percolation transition we study 
   the percolating system when the OP undergoes its fastest rate of growth. As links are being dropped at unit rate we keep track of 
   $s_{max}$ and calculate how much $s_{max}$ increases due to each link addition. In a certain run, let the maximal jump in $s_{max}$ 
   occur due to addition of $n_{max}$-th link. We then define the percolation threshold of this specific run as $p_{max}(N)=n_{max}/N$.
   Consequently the percolation threshold $p_c(N)$ of a graph of size $N$ has been determined by the average of $p_{max}(N)$ values over 
   a large number of runs, i.e., 
\begin {equation}
p_c(N) = \langle p_{max}(N) \rangle.
\end {equation} 

\begin{figure}
\begin {center}
\includegraphics[width=7.0cm]{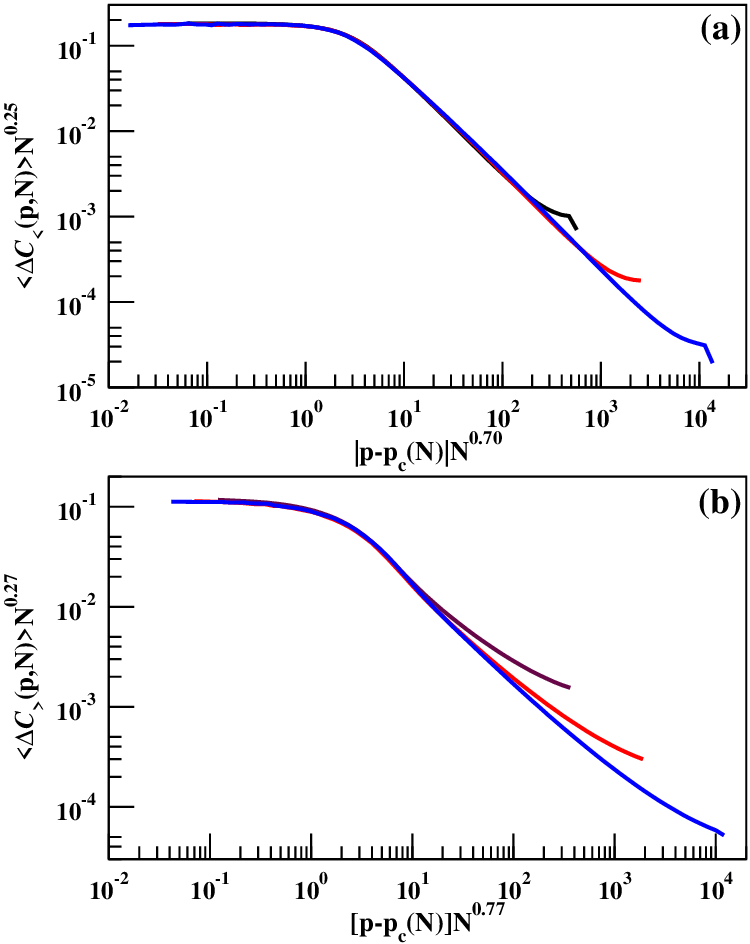}
\end {center}
\caption{(Color online) For AP the finite-size scaling of average size of the jump in OP for $N=10^4$ (blue), $10^5$ (red) and $10^6$ (green).
(a) Sub-critical region. The average jump size $\langle \Delta {\cal C}_<(p,N) \rangle$ is plotted with deviation $|p-p_c(N)|$.
The average slopes give $\alpha_<$ = 1.18(5).
(b) Super-critical region. The average jump size $\langle \Delta {\cal C}_>(p,N) \rangle$ is plotted with deviation $p-p_c(N)$.
The average slopes give $\alpha_>$ = 1.02(5).
}
\end{figure}

      It is first noted that there is hardly any unique largest component below the percolation threshold. More specifically addition 
   of a single link may merge two smaller components whose combined size may be larger than $s_{max}$. To get a first hand idea we plot 
   in Fig. 1(a) using blue symbols the successive jump heights in OP i.e., $\Delta {\cal C}(N)$ vs. $p$ in a single run for AP with 
   $N=10^6$. Further we re-plot a subset of these points by red symbols for whom the jumps in $s_{max}$ occurred to the already 
   existing largest component. This plot shows that the percolation threshold $p_{max}(N)$ clearly distinguishes two regimes. In the
   sub-critical regime $p < p_{max}(N)$ there is few red points, signifying a unique largest component of the graph does not 
   exist in this regime of link density. On the other hand in the super-critical regime $p > p_{max}(N)$ the jumps occur much 
   more frequently than in the sub-critical regime since the largest component is already so big that almost any link addition enhances 
   its size. There cannot be a blue point in this regime since that would imply the appearance of another largest 
   component, which will eventually merge with present one with an even larger jump in ${\cal C}$ which is not possible since the
   maximal jump in ${\cal C}$ has already occurred. 

      In Fig. 1(b) we plot the average jump size $\langle \Delta {\cal C}(p,N) \rangle$ 
   with the deviation $p-p_c(N)$ from the percolation threshold. The variation of $\Delta {\cal C}(p,N)$ is different in the
   sub-critical regime and in the super-critical regime. This looks similar to the well known $\lambda$-transition of the divergence of 
   specific heat in critical phenomena characterized by the same value of exponent but with different amplitudes for the sub and 
   super-critical regimes. In comparison we get different exponents as reported below for the two sides of the percolation threshold.

\begin{figure}[top]
\begin {center}
\includegraphics[width=7.0cm]{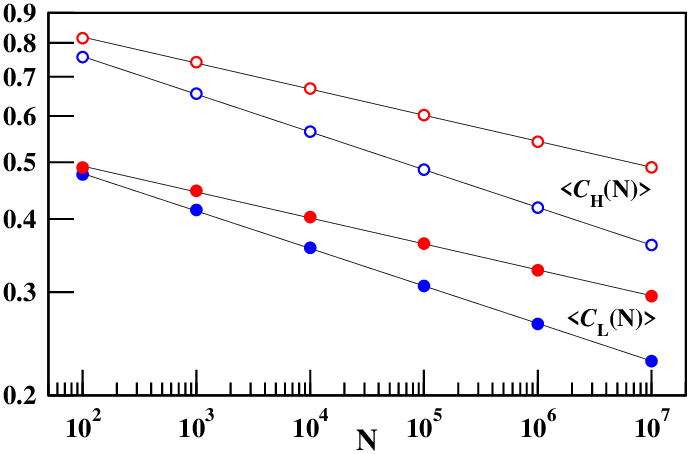}
\end {center}
\caption{(Color online) The average values of $\langle {\cal C}_L(N) \rangle$ (filled circles) and $\langle {\cal C}_H(N) \rangle$ 
(empty circles) are plotted with $N$ for AP (blue) and for da Costa model (red). From slopes the growth exponent $\chi$ of the largest 
component are estimated as 0.9355(5) for AP and 0.9553(5) for da Costa model.
}
\end{figure}

      In Fig. 2 this analysis has been done in more detail and a finite-size scaling analysis of $\langle \Delta {\cal C}(p,N) \rangle$
   is shown. In the sub-critical regime the average jump size $\langle \Delta C_<(p,N) \rangle$ have been calculated as a function of
   $|p-p_c(N)|$. For all three system sizes the plots of binned data on a $\log - \log$ scale exhibit similar 
   behavior, i.e., an initial horizontal part for very small values of $|p-p_c(N)|$ followed by a linear downward regime, indicating a 
   power law decay like:
\begin {equation}
\langle \Delta C_<(p,N) \rangle \sim |p-p_c(N)|^{-\alpha_<}.
\end {equation}
   The directly measured values of $\alpha_<$ are 1.133, 1.157 and 1.164 respectively for the three system sizes which approach to 1.18(5)
   for large $N$. In Fig. 2(a) a finite-size scaling analysis of this data has been done using the form:
\begin {equation}
\langle \Delta C_<(p,N) \rangle N^{0.25} \sim {\cal G}_<[|p-p_c(N)|N^{0.70}].
\end {equation}
   For the super-critical regime the average jump size $\langle \Delta C_>(p,N) \rangle$ in OP has been plotted with $p-p_c(N)$ in Fig. 2(b). 
   A similar variation of initial flat, followed by a power law decay $\langle \Delta C_>(p,N) \rangle \sim (p-p_c(N))^{-\alpha_>}$ has been
   found. The directly measured values of $\alpha_>$ are 0.79, 0.89 and 0.92 respectively which approach to 1.02(5) for large $N$. A similar 
   data collapse was possible using the following scaling equation: 
\begin {equation}
\langle \Delta C_>(p,N) \rangle N^{0.27} \sim {\cal G}_>[(p-p_c(N))N^{0.77}].
\end {equation}

\begin{figure}[top]
\begin {center}
\includegraphics[width=7.0cm]{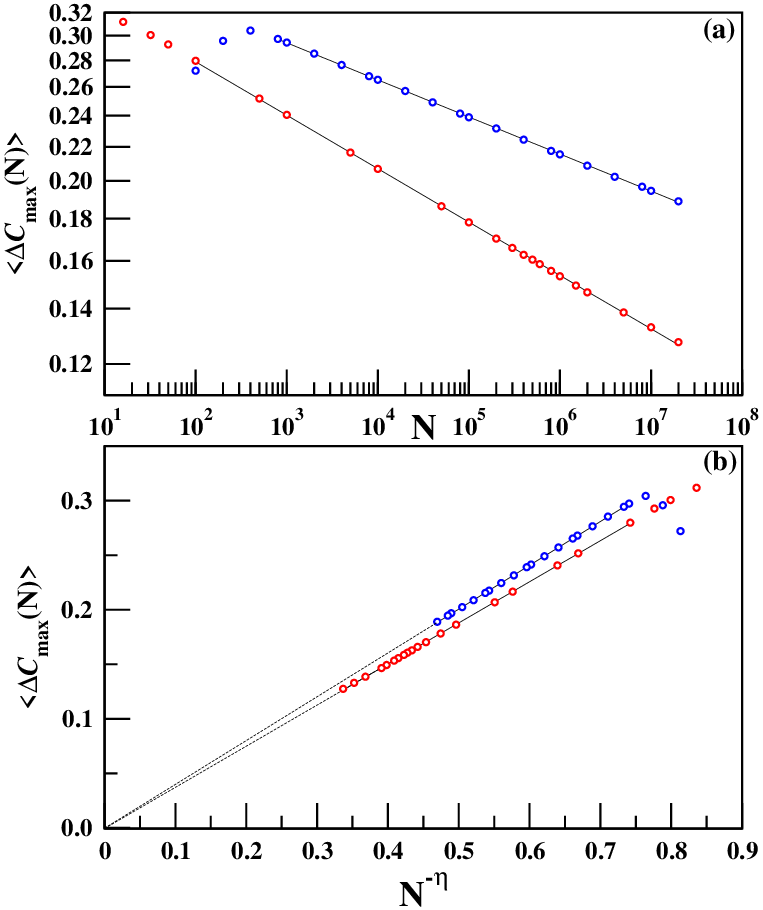}
\end {center}
\caption{(Color online)
(a) The average maximal jump $\langle \Delta {\cal C}_{max}(N) \rangle$ in OP plotted on a $\log-\log$ scale with graph size $N$. 
The slopes are $\eta$(AP)$\approx 0.06475$ (red) and $\eta$(da Costa)$\approx 0.04495$ (blue).
(b) The same sets of data are plotted against $N^{-\eta}$ on a lin-lin scale. Continuous lines are straight line fits
    of the data which are then extrapolated to $N \to \infty$ to meet $\langle \Delta {\cal C}_{max}(N) \rangle$ axis at 
    $\langle \Delta {\cal C}_{max}(\infty) \rangle \approx -0.000516$ for AP (red) and
    $\approx -0.000256$ for da Costa model (blue).}
\end{figure}

\begin{figure*}[top]
\begin {center}
\includegraphics[width=15.0cm]{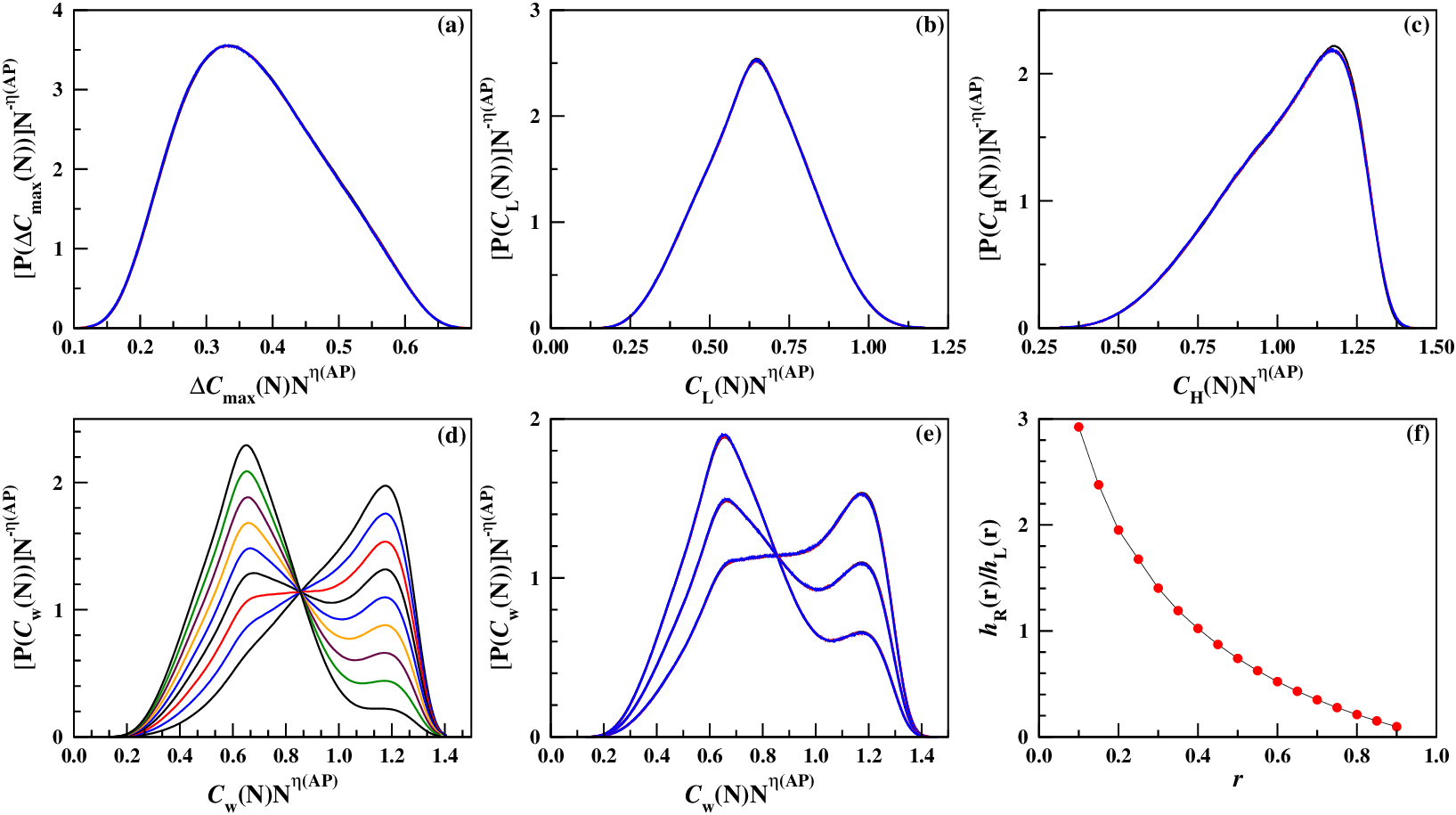}
\end {center}
\caption{(Color online) For Achlioptas Process on graphs of sizes $N = 10^3$ (black), $10^4$ (red) and $10^5$ (blue). 
Finite-size scaling analysis are displayed for different probability distributions associated with the Order
Parameter ${\cal C}$ using $\eta$(AP)$=0.0645$. They are:
(a) Maximal jump sizes $\Delta {\cal C}_{max}(N)$.
(b) The lower end value ${\cal C}_L(N)$ of the maximal jump in ${\cal C}$.
(c) The higher end value ${\cal C}_H(N)$ of the maximal jump in ${\cal C}$.
(d) The interpolating function ${\cal C}_w(N)$ of ${\cal C}_L(N)$ and ${\cal C}_H(N)$ with probabilities $r$ and $1-r$. The right-top 
curve is for $r$=0.1 and increased to 0.9 at the interval of 0.1 using only one system size $N=10^3$. 
(e) The scaling of $P({\cal C}_w(N))$ for $r$ = 0.3 (right-top), 0.5 and 0.7.
(f) The ratio of heights of the right peak $h_R(r)$ and left peak $h_L(r)$ of the bimodal distributions in 5(d) plotted with 
probability $r$.
}
\end{figure*}

      It seems likely that growth rules of few leading components are similar at the percolation threshold. For example the
   largest component, second largest component etc. may grow in the same way since they grow `independently' but in very 
   similar environments. We define the `growth exponent' $\chi$ for the size of the largest component (and next few leading 
   components) at the percolation threshold as $\langle s_{max}(p_c(N),N) \rangle \sim N^{\chi}$ or in terms of OP 
   $\langle {\cal C}(p_c(N),N) \rangle \sim N^{-\eta}$ with $\eta=1-\chi$. Since for regular $d$-dimensional 
   lattices $N=L^d$, percolation on these lattices has the growth exponent $\chi = D/d$ where $D$ is the fractal dimension of 
   the `infinite' incipient cluster at the percolation threshold.

      While the size of the largest component monotonically increases with link density size of the second largest component
   gradually increases, reaches a maximum and then suddenly falls to a small value \cite {Lee}. The maximal jump in the OP 
   occurs precisely at this point when the largest component merges with the largest (in the entire history of the system)
   second largest component. Since both of 
   them grow in a similar way it is expected that the height of the maximal jump in the largest component should also grow
   as $N^{\chi}$. Let us assume that during the fastest growth OP jumps from ${\cal C}_L(N)$ to ${\cal C}_H(N)$ so that 
   $\Delta {\cal C}_{max}(N)= {\cal C}_H(N) - {\cal C}_L(N)$. In Fig. 3 we plot on a $\log - \log$ scale the average values 
   of $\langle {\cal C}_L(N) \rangle$ and $\langle {\cal C}_H(N) \rangle$ with $N$ for AP. Linear least square fits exhibit 
   that both plots are highly straight, no systematic variation of slopes could be detected and they are very closely parallel 
   to each other with slopes -0.0641(5) and -0.0645(5) giving an average of $\eta$(AP)=0.0643(5) and the growth 
   exponent $\chi$(AP)$ = 0.9357(5)$. Similar data for the da Costa model have also been plotted in Fig. 3 with red color 
   and the estimates for the slopes are -0.0443(5) and -0.0445(5) corresponding to the growth exponent of 
   $\chi$(da Costa)$ = 0.9556(5)$.

\begin{figure}
\begin {center}
\includegraphics[width=7.0cm]{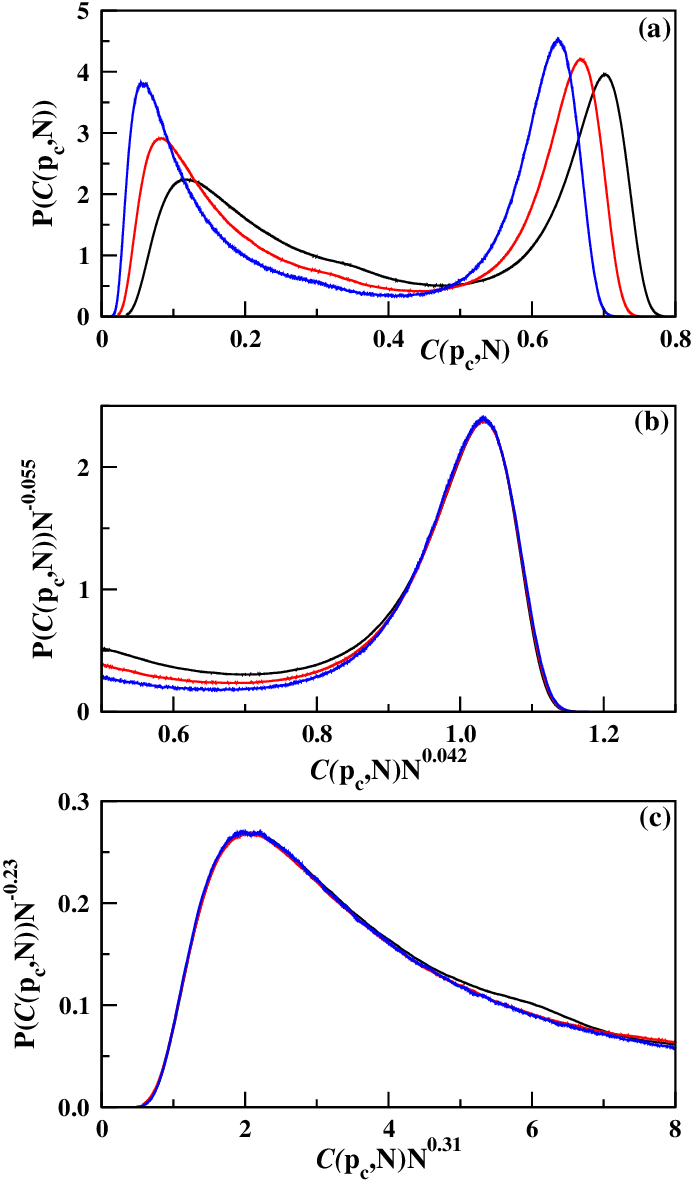}
\end {center}
\caption{(Color online) Distribution of the Order Parameter ${\cal C}(p_c,N)$ right at the percolation threshold $p_c(N)$. 
(a) $P({\cal C}(p_c,N))$ plotted with ${\cal C}(p_c,N)$ for $N$ = 10000 (black), 32000 (red) and 100000 (blue).
Finite-size scaling of the right peak in (b) and that of the left peak in (c).}
\end{figure}

      In addition, size of the maximal jump has also been estimated directly and its average value for different 
   system sizes are then extrapolated as:
\begin {equation}
\langle \Delta {\cal C}_{max}(N) \rangle = \langle \Delta {\cal C}_{max}(\infty) \rangle + AN^{-\eta}
\end {equation}
   as $N \to \infty$ and if $\langle \Delta {\cal C}_{max}(\infty) \rangle$ turns out to be zero or non-zero it 
   signifies a continuous / discontinuous change in OP. In Fig. 4(a) we plot $\langle \Delta {\cal C}_{max}(N) \rangle$ 
   vs. $N$ for AP using a $\log - \log$ scale. A straight line fits to the data very accurately except for very small 
   system sizes and gives $\eta$(AP)$ = 0.06475$. In Fig. 4(b) we use this value of $\eta$ and plot $\langle \Delta {\cal C}_{max}(N) \rangle$
   vs. $N^{-\eta}$ on a lin - lin scale. Here again the least square fit of a straight line works nicely, giving 
   $\langle \Delta {\cal C}_{max}(\infty) \rangle = -0.000516$. This analysis suggests that indeed the form of Eqn. (4) 
   is very likely to be valid and since the constant $\langle \Delta {\cal C}_{max}(\infty) \rangle$ is nearly equal to zero, it 
   gives a strong indication that the transition of AP is actually continuous. Similarly for da Costa 
   model we get $\eta$(da Costa)$ = 0.04495$ and $\langle \Delta {\cal C}_{max}(\infty) \rangle = -0.000256$. We conclude
   $\eta$(AP)= 0.0645(5), $\chi$(AP)$ = 0.9355(5)$; $\eta$(da Costa)= 0.0447(5), $\chi$(da Costa)$ =  0.9553(5)$
   and use these values in the following analysis.

      Fig. 5 shows the plots for the finite-size scaling analysis of a number of probability distributions associated with the 
   maximal jump of OP in the Achlioptas Process. They are for: (a) the maximal jump size $\Delta {\cal C}_{max}(N)$; (b) the lower 
   end value ${\cal C}_L(N)$ of the maximal jump; (c) the higher end value ${\cal C}_H(N)$ of the maximal jump.
   While the distribution $P({\cal C}_L(N))$ appears to be almost symmetric the distribution $P({\cal C}_H(N))$ is found to be 
   quite asymmetric. For the scaling analysis three system sizes are used in each case and the data collapse are found to be 
   excellent. In each case the $x$-axis is scaled by $N^{\eta(AP)}$ and the $y$-axis is scaled by $N^{-\eta(AP)}$. We further 
   define a single value ${\cal C}_w$ of the Order Parameter at the maximal jump as ${\cal C}_w = {\cal C}_L$ with probability 
   $r$ and ${\cal C}_w = {\cal C}_H$ with probability $1-r$. One can define the interpolating function with respect to the
   common variable ${\cal C}$ as: $P_r({\cal C}) = r P_L({\cal C}) + (1-r) P_H({\cal C})$. In Fig. 5(d) we show this interpolating
   function $P({\cal C}_w(N))$ having double humps with unequal heights and widths for 9 values of $r$ from 0.1 to 0.9 at an 
   interval of 0.1 for only one size $N$=1000. In Fig. 5(e) we plot the scaling analysis of $P({\cal C}_w(N))$ for few representative 
   values of $r$ = 0.3, 0.5 and 0.7 which also scale with $N^{\eta(AP)}$. In both Fig. 5(d) and 5(e) the common point of intersection
   where all curves pass through has coordinates (0.855, 1.14). Finally in Fig. 5(f) we plot the ratio $h_R(r)/h_L(r)$ of the heights 
   of right peak and left peak as a function of $r$. Therefore it turns out that only one exponent $\eta$(AP) which determines 
   the growth exponent $\chi$(AP)$ = 1-\eta$(AP) characterizes various distributions related to the fastest growth of the Order 
   Parameter in AP. In addition it is also verified that for da Costa model all these probability distributions are very much 
   similar and they scale equally well with the corresponding value of $\eta$(da Costa).

\begin{figure}
\begin {center}
\includegraphics[width=7.0cm]{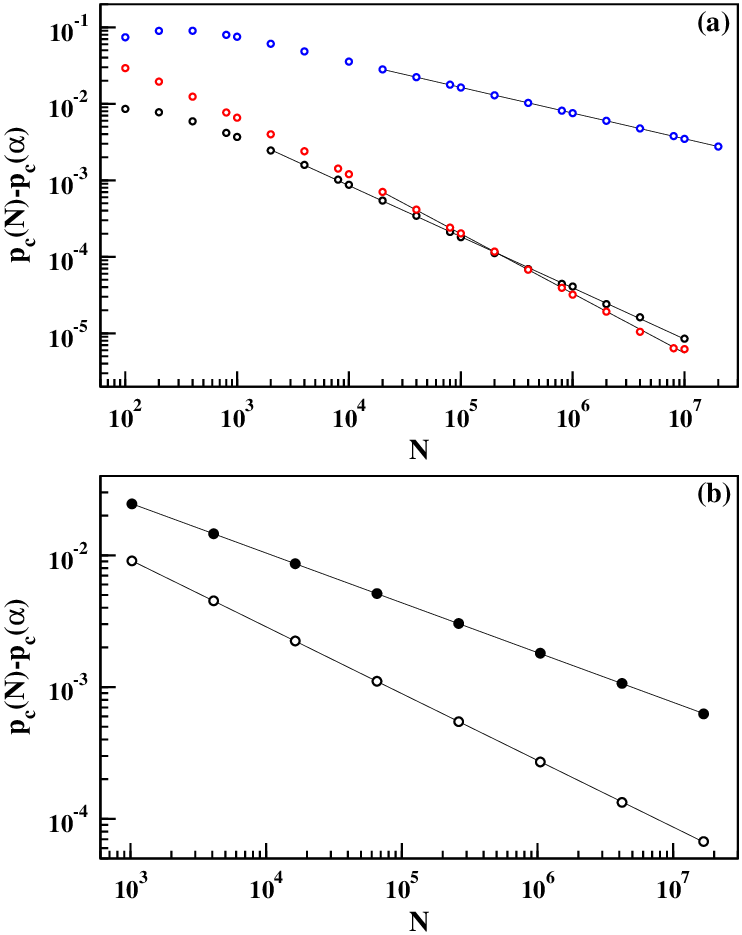}
\end {center}
\caption{(Color online) Plot of $p_c(N) - p_c(\infty)$ vs. $N$ on the $\log - \log$ scale for AP (black), da Costa model (red) and Random Graphs
(blue). (a) Apart from few small size graphs the data points fit nicely to straight lines giving values of the $\nu$ exponent
in Eqn. (6) as 1.50, 1.26 and 2.98 respectively. 
(b) Similar data for the AP (empty circles) and ordinary bond percolation (filled circles) on square lattice giving $\nu$ = 1.98 and
2.67 respectively.}
\end{figure}

      Bimodal distributions have been observed for the probability of the Order Parameter at the percolation threshold 
   \cite {Grassberger,Radicchi1,Tian}. However this distribution depends on the precise definition of the percolation
   threshold. In \cite {Grassberger} the percolation threshold $p'_c(N)$ has been determined in such a way so that heights
   of the two peaks are equal and this is kept fixed for all runs. In comparison in our case different runs have different
   percolation thresholds $p_{max}(N)$ since the maximal jumps $\Delta {\cal C}_{max}(N)$ in OP occur at these specific link densities.
   These are two different statistical ensembles though they approach the same percolation point $p_c(\infty)$ as $N \to \infty$. 
   Evidently in a single run $p_{max}(N)$ may be either smaller or larger than $p'_c(N)$. If $p'_c(N) < p_{max}(N)$ then 
   its corresponding OP ${\cal C} < {\cal C}_L$ and when $p'_c(N) > p_{max}(N)$ then its ${\cal C} > {\cal C}_H$. For this 
   reason with a fixed run-independent value of $p'_c(N)$ there are two regions where ${\cal C}$ occur more frequently than 
   its other values resulting a bimodal distribution of OP. The peaks of the bimodal distribution occurring at 
   ${\cal C}_{\pm}$ with ${\cal C}_+ > {\cal C}_-$ scale as ${\cal C}_{\pm} \sim N^{-\eta_\pm}$ \cite {Grassberger}. We observe
   that our exponent $\eta$ lies in between them i.e., $\eta_+ < \eta < \eta_-$ for both AP and da Costa model (Table I). 

      Further we calculate the distribution of the Order Parameter at $p_c(N)$ defined in Eqn. (1) and fixed for all runs.
   A sample size of $10^7$ has been used for three different system sizes $N$ = 10000, 32000 and 100000. These are bimodal
   distributions $P({\cal C}_{p_c}(N))$ with unequal heights and widths shown in Fig. 6(a). We see positions of both peaks 
   ${\cal C}_{\pm}(p_c,N)$ decreases as $N$ increases. The finite-size scaling of the right and left peaks are displayed in 
   Fig. 6(b) and Fig. 6(c). Assuming the same scaling forms ${\cal C}_{\pm}(p_c,N) \sim N^{-\eta_\pm}$ we obtain $\eta_+ = 
   0.042$ and $\eta_-=0.31$. Here also the same inequality $\eta_+ < \eta < \eta_-$ holds good. 

\begin{figure}
\begin {center}
\includegraphics[width=7.0cm]{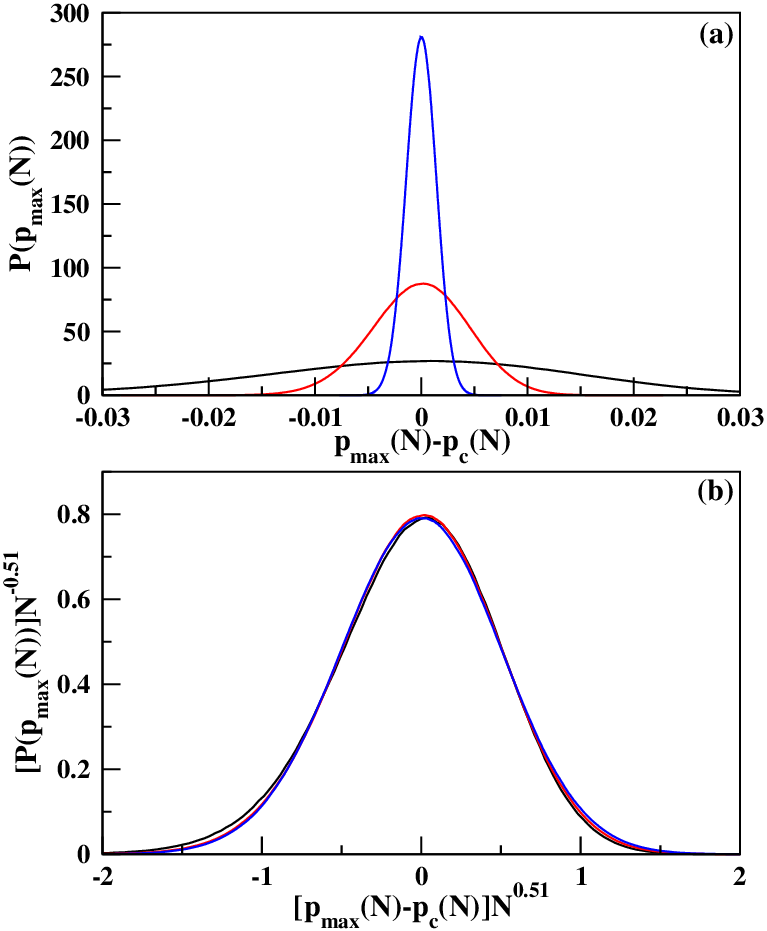}
\end {center}
\caption{(Color online) For AP the probability distributions $P(p_{max}(N))$ vs. $p_{max}(N)-p_c(N)$ for three system sizes 
$N = 10^3$ (black), $10^4$ (red) and $10^5$ (blue) in (a) and their finite-size scaling in (b).}
\end{figure}

      Next we make a systematic study how the percolation threshold $p_c(N)$ approaches to its asymptotic value $p_c(\infty)$.
   We cannot talk about a correlation length exponent in this case since graphs are not embedded in Euclidean space, yet
   we assume a leading order correction for the finite size systems associated with an exponent $\nu$ (not the correlation
   length exponent) as:
   \begin {equation}
   p_c(N) = p_c(\infty) + BN^{-1/\nu}.
   \end {equation}
   In Fig. 7(a) we plot $p_c(N) - p_c(\infty)$ vs. $N$ on a $\log - \log$ scale using $p_c(\infty) = 0.888446$ (note that this differs
   3 in the last digit from the value quoted in \cite {Grassberger}) for AP, 0.923207508 for the da Costa model \cite {Costa} and 1/2 
   for the Random Graphs \cite {Erdos}. On such a scale each plot is expected to be linear if the corresponding $p_c(N)$ values obey 
   the Eqn. (6). It is observed that apart from very small graphs the data points do fit to straight lines. The slopes in these regions 
   give $\nu$ values as 1.50, 1.26 and 2.98 for AP, da Costa model and Random Graphs which we conjecture may be 3/2, 5/4 and 3 exactly.
   We further plot similar data of AP on the square lattice in Fig. 7(b) using $p_c(\infty) = 0.526565$ \cite {ZIFF1}.
   The curve is a very nice straight line with slope -0.505 corresponding to $\nu \approx 1.98$ which leads us to conjecture $\nu_{AP}=2$
   exactly for square lattice. With this result it seems to be better to fit $p_c(N)$ data for square lattice of size $L$ (with 
   $N=L^2$) as $p_c(N) = p_c(\infty) + c_1N^{-1/2}+c_2N^{-1}$ \cite {ZIFF2}. Using this quadratic fitting form and with eight 
   data points from $L=32$ to $4096$ we get $p_c(\infty) = 0.526575$. The extrapolated value systematically came down slowly
   when we discarded the data for small lattices. Finally for the last three points $L=$ 1024, 2048 and 4096 we get $p_c(\infty) 
   = 0.5265639$. In addition we repeat this calculation for ordinary bond percolation on square lattice and using $p_c(\infty)=1/2$ 
   we get a good straight line with $\nu \approx 2.67$ (Fig. 7(b)). 

\begin{figure}
\begin {center}
\includegraphics[width=7.0cm]{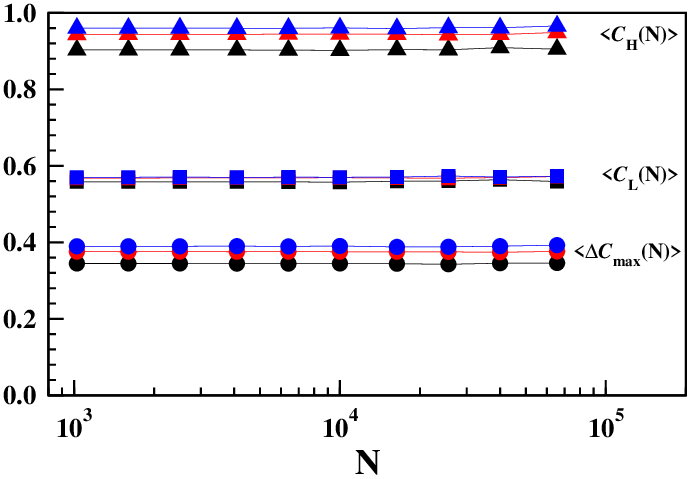}
\end {center}
\caption{(Color online) For the Global rule EP model \cite {Manna} on square lattice the values of 
$\langle {\cal C}_L(N) \rangle$ (square), $\langle {\cal C}_H(N) \rangle$ (triangle) and $\langle \Delta {\cal C}_{max}(N) \rangle$
(circle) are plotted for $\zeta$ = -1 (black), -2(red) and -3(blue).}
\end{figure}

      Next we calculate for AP the probability density distribution of the percolation thresholds $P(p_{max}(N))$ and plot them with 
   $p_{max}(N) - p_c(N)$ in Fig. 8(a) again for the same three different system sizes and with $10^7$ configurations for each system 
   size. All three plots fit very closely to Gaussian distributions with different parameter values for different $N$. Fig. 8(b) shows 
   the finite-size scaling of these data and the best data collapse is obtained corresponding to the following form:
\begin {equation}
P(p_{max}(N)) \sim N^{\theta}{\cal F}([p_{max}(N)-p_c(N)]N^{\theta}).
\end {equation}
   A best value of $\theta=0.51$ is obtained which is very much consistent with the conjecture of $\theta=1/2$ in \cite {Grassberger}.

      Finally we calculate the growth exponent $\chi$ again for the Global rule EP model \cite {Manna}. 
   It was argued that this model exhibits discontinuous transitions for all values of the biasing parameter $\zeta < 0$ on the 
   square lattice. We therefore calculate $\langle {\cal C}_L(N) \rangle$, $\langle {\cal C}_H(N) \rangle$ and 
   $\langle \Delta {\cal C}_{max}(N) \rangle$ for $\zeta = -1, -2$ and -3. In Fig. 9 we plot them with $N$ on a lin-log scale
   and see that curves are simply horizontal straight lines indicating $N$ independence. This implies that the growth exponent $\chi=1$ and
   $\eta=0$ for the Global rule EP model \cite {Manna}. Therefore in general $\langle \Delta {\cal C}_{max}(N) \rangle$ = constant
   implies that for these $\zeta$ values the transitions are discontinuous.

\begin{table}[top]
\begin {center}
\begin{tabular}{|c|r|r|r|r|c|} \hline
        & AP                              & da Costa   & AP in 2$d$ & BP  & RG    \\ \hline
$p_c$   & 0.888449(2) \cite {Grassberger} & 0.923207508 \cite {Costa} & 0.526565 \cite {ZIFF1}          & 1/2    & 1/2     \\
        & 0.888446                        &                           & 0.5265639                       &        &         \\
$\eta_+$& 0.0402(15) \cite {Grassberger}  & 0.0255(80) \cite {Grassberger} & 0.018(2) \cite {Grassberger}&        &\\
$\eta$  & 0.0645(5)                       & 0.0447(5)                    & 0.0217(5)                     & 5/96 & 1/3 \\
$\eta_-$& 0.270(7) \cite {Grassberger}    & 0.300(5)  \cite {Grassberger} & 0.078(7) \cite {Grassberger} &        & \\
$\chi$  & 0.9355(5)                       & 0.9553                    & 0.9783(5)                       & 91/96 & 2/3 \\ 
$\nu$   & 1.50                            & 1.26                      & 1.98                           & 8/3   & 2.98 \\ \hline
\end{tabular}
\caption{Values of different exponents available in the literature as well as measured in this work. Some known
results of Bond Percolation (BP) in 2$d$ and for Random Graphs are also included for comparison. The $p_c$, $\eta_+$ and $\eta_-$ 
values are taken form \cite {Grassberger}.}
\end {center}
\end {table}

      We summarize that by looking at the numerical value of the exponent $\chi$ describing the growth of the largest component at 
   the percolation thresholdwith with graph size $N$ one should be able to understand if the model exhibits a continuous or discontinuous
   transition. The growth exponent is defined as $\langle s_{max} \rangle \sim N^{\chi}$ has been quite well known for graphs in the 
   literature but perhaps without a name. This exponent is similar to the fractal dimension when the graph is embedded in Euclidean space
   i.e., $\chi < 1$ implies a `fractal' with fractal dimension being less than the embedding Euclidean space dimension and $\chi=1$ implies 
   a `compact' giant component whose size grows proportional to the graph size $N$. We justify this as follows. The maximal jump in the Order
   Parameter takes place only when the largest component merges with the largest second largest component and sizes of both grow as $N^{\chi}$
   at the percolation threshold. For a continuous transition the maximal jump in OP must have a zero measure with respect to 
   the graph size $N$ in the asymptotic limit and therefore $\chi < 1$. On the other hand for a discontinuous transition, the largest 
   jump in OP which is the size of the largest second largest component must be proportional to the graph size $N$ and therefore $\chi=1$. 
   We conclude that any percolation model which has $\chi < 1$ / $\chi = 1$ must exhibit continuous / discontinuous transition and the 
   reverse is also true. In addition we show that a number of distributions related to the largest jump of OP at the percolation threshold 
   are described by a related exponent $\eta=1-\chi$. A comparison of all related exponents have been done in Table 1.

      In recent times it is being shown that asymptotically all Local rule models have continuous transitions \cite
   {Riordan,Grassberger,Costa,Manna,Nagler,Tian}. In comparison Global rule models exhibit discontinuous transitions. We conjecture 
   that for all Local rule models with $\chi < 1$ transitions are continuous, whereas all Global rule models have 
   $\chi = 1$ and their transitions are discontinuous.

      Finally we would like to make the following comment. Melting of ice is a well known example of first order transition where
   the density of ice changes from 0.92 gm/ml to 1.00 gm/ml of water. On the other hand on a graph with Avogadro number of nodes
   ($N \approx 10^{23}$) the Achlioptas process has the largest jump in the Order Parmeter as $\Delta {\cal C}_{max}(N) = N^{-\eta(AP)} 
   \approx 0.03$ with $\eta(AP)=0.0645$. Therefore we believe that `practically' the transition in Achlioptas process may very well 
   be considered as the discontinuous transition. 
   
      I am very much indebted to Deepak Dhar, Peter Grassberger, Maya Paczuski and Robert M. Ziff for many helpful suggestions. 
   I convey my sincere thanks to Arnab Chatterjee, Raissa D'Souza and Janos Kertesz for the critical reading of the manuscript.

\leftline {manna@bose.res.in}


\begin{thebibliography}{90}
\bibitem {Stanley} H. E. Stanley, {\it Introduction To Phase Transitions And Critical Phenomena}, Oxford University Press, 1971.
\bibitem {Stauffer} D. Stauffer and A. Aharony, {\it Introduction to Percolation Theory} (Taylor \& Francis, London, 1994).
\bibitem {Grimmett} G. Grimmett, {\it Percolation}, Springer, 1999.
\bibitem {Broadbent} S. Broadbent, J. Hammersley,  {\it Percolation processes I. Crystals and mazes}, Proceedings of the Cambridge Philosophical Society {\bf 53}, 629 (1957).
\bibitem {ACH} D. Achlioptas, R. M. D'Souza, and J. Spencer, Science {\bf 323}, 1453 (2009).
\bibitem {Erdos} P. Erd\H os and A. R\'enyi, Publ. Math. Debrecen, {\bf 6}, 290 (1959).
\bibitem {Grassberger} P. Grassberger, C. Christensen, G. Bizhani, S-W. Son and M. Paczuski, Phys. Rev. Lett. {\bf 106}, 225701 (2011).
\bibitem {ZIFF} R. M. Ziff, Phys. Rev. Lett. {\bf 103}, 045701 (2009).
\bibitem {ZIFF1} R. M. Ziff, Phys. Rev. E {\bf 82}, 051105 (2010).
\bibitem {Friedman} E. J. Friedman A. S. Landsberg, Phys. Rev. Lett. {\bf 103}, 255701 (2009).
\bibitem {Cho} Y.S. Cho, J. S. Kim, J. Park, B. Kahng and D. Kim, Phys. Rev. Lett. {\bf 103}, 135702 (2009).
\bibitem {Radicchi} F. Radicchi and S. Fortunato, Phys. Rev. Lett. {\bf 103}, 168701 (2009).
\bibitem {Radicchi1} F. Radicchi and S. Fortunato, Phys. Rev. E, {\bf 81}, 036110 (2010).
\bibitem {Herrmann} A. A.Moreira, E. A. Oliveira, S. D. S. Reis, H. J. Herrmann and J. S. Andrade Jr. Phys. Rev. E {\bf 81}, 040101 (2010);
N. A. M. Ara\'ujo and H. J. Herrmann, Phys. Rev. Lett. {\bf 105}, 035701 (2010).
\bibitem {Cho1} Y. S. Cho, B. Kahng and D. Kim, Phys. Rev. E {\bf 81}, 030103(R) (2010).
\bibitem {Rozenfeld} H. D. Rozenfeld, L. K. Gallos and H. A. Makse, Eur. Phys. J. B {\bf 75}, 305 (2010).
\bibitem {Pan} R. K. Pan, M. Kivel\"a, J. Sarama\"ki, K. Kaski, J. Kert\'esz, Phys. Rev. E {\bf 83}, 046112 (2011).
\bibitem {Costa} R. A. da Costa, S. N. Dorogovtsev, A. V. Goltsev and J. F. F. Mendes, Phys. Rev. Lett. {\bf 105}, 255701 (2010).
\bibitem {Manna} S. S. Manna and A. Chatterjee, Physica A, {\bf 390}, 177 (2011).
\bibitem {Nagler} J. Nagler, A. Levina and M. Timme, Nature Physics {\bf 7}, 265 (2011).
\bibitem {Riordan} O. Riordan and L. Warnke, arxiv: 1102.5306.
\bibitem {Tian} L. Tian and D-N. Shi, arXiv:1010.5990.
\bibitem {Lee} H. K. Lee, B. J. Kim and H. Park, arxiv: 1103.4439.
\bibitem {ZIFF2} R. M. Ziff, private communication.
\end {thebibliography}
\end {document}